# The Core of NGC 6240

# from Keck Adaptive Optics and HST NICMOS Observations

C. E. Max[1], G. Canalizo[2], B. A. Macintosh[3], L. Raschke[4],

D. Whysong[5], R. Antonucci[5], G. Schneider[6]

[1] Center for Adaptive Optics, University of California, 1156 High Street, Santa Cruz CA 95064 and Institute of Geophysics and Planetary Physics, Lawrence Livermore National Laboratory, 7000 East Avenue, Livermore CA 94550, max@ucolick.org and max1@llnl.gov

[2] Current address: Department of Earth Sciences and Institute of Geophysics and Planetary Physics, University of California, Riverside CA 92521, gabriela.canalizo@ucr.edu

[3] I Division, Lawrence Livermore National Laboratory, 7000 East Avenue, Livermore CA 94550, bmac@igpp.ucllnl.org

[4] Center for Adaptive Optics, University of California at Santa Cruz, 1156 High Street, Santa Cruz CA 95064, lynne@ucolick.org

[5] Physics Department, University of California at Santa Barbara, Santa Barbara CA 93106, ski@spot.physics.ucsb.edu and dwhysong@physics.ucsb.edu

[6] Steward Observatory, University of Arizona, Tucson, AZ 85721, gschneider@stsci.edu

ABSTRACT

We present results of near infrared imaging of the disk-galaxy-merger NGC 6240 using adaptive optics on the Keck II Telescope and reprocessed archival data from NICMOS on the Hubble Space Telescope. Both the North and South nuclei of NGC 6240 are clearly elongated, with considerable sub-structure within each nucleus. In K' band there are at least two point-sources within the North nucleus; we tentatively identify the south-western point-source within the North nucleus as the position of one of the two AGNs. Within the South nucleus, the northern sub-nucleus is more highly reddened. Based upon the nuclear separation measured at 5 GHz, we suggest that the AGN in the South nucleus is still enshrouded in dust at K' band, and is located slightly to the north of the brightest point in K' band. Within the South nucleus there is strong $H_2$ 1-0 S(1) line emission from the northern sub-nucleus, contrary to the conclusions of previous seeing-limited observations. Narrowband $H_2$ emission-line images show that a streamer or ribbon of excited molecular hydrogen connects the North and South nuclei. We suggest that this linear feature corresponds to a bridge of gas connecting the two nuclei, as seen in computer simulations of mergers. Many point-like regions are seen around the two nuclei. These are most prominent at 1.1 microns with NICMOS, and in K'-band with Keck adaptive optics. We suggest that these point-sources represent young star clusters formed in the course of the merger.

*Subject headings*: galaxies: active, galaxies: interactions, galaxies: individual (NGC 6240), instrumentation: adaptive optics

1. INTRODUCTION

Mergers and interactions of gas-rich galaxies are thought to be important triggers both for nuclear activity and for starbursts (e.g. Kennicutt and Keel 1984; Norman & Scoville 1988; Sanders & Mirabel 1996; Barnes & Hernquist 1996; Genzel et al. 1998). Indeed the resulting starbursts and active galactic nuclei (AGNs) may well be causally related. Winds and outflows from starbursts can provide fuel to "feed" the AGNs' central black hole (e.g. Canalizo & Stockton, 2001), and gravitational torques can remove angular momentum from shocked galactic gas (Barnes & Hernquist 1991, 1996) allowing the gas to fall in toward the black hole and fuel it.

NGC 6240 (z = 0.0243, d = 98 Mpc for $H_0$ = 75 km s$^{-1}$ Mpc$^{-1}$, 1 arc sec = 470 pc) is a merger of two massive disk galaxies (Fosbury & Wall 1979; Tacconi et al. 1999). It contains two nuclei separated by 1.5 – 1.7 arc sec (see for example optical images in Gerssen et al. 2004; Rafanelli et al. 1997; near-IR images in Scoville et al. 2000; radio images in Beswick et al. 2001 and in Gallimore & Beswick 2004). In the optical through infrared, the angular separation between the two nuclei is known to decrease toward longer observing wavelength, suggesting heavy dust extinction in the nuclear regions. On larger spatial scales, the optical emission of NGC 6240 shows a dramatic "bow-tie" structure with tidal tails of the type seen in computer simulations of merging disk galaxies (e.g. Toomre & Toomre 1972) and observed in many other galaxy mergers. NGC 6240 has been known from x-ray observations by Bepposax (Vignati et al. 1999) and ASCA (Iwasawa & Comastri 1998) to contain at least one AGN, highly absorbed at x-ray wavelengths. In November 2002 it was announced (Komossa et al. 2003) that CHANDRA high-energy x-ray observations using the ACIS-S detector have resolved two AGNs, one in each nucleus, with approximately the same separation as the two compact nuclei seen at 5 GHz by MERLIN (Gallimore & Beswick 2004). Visible-light spectroscopy classifies NGC 6240 as a LINER (Veilleux et al. 1995).

Infrared observations of NGC 6240 by IRAS (Wright et al. 1984) and ISO (Genzel et al. 1998; Lutz et al. 2003) classify it as a Luminous Infrared Galaxy (LIRG) and show that NGC 6240 contains an ongoing starburst. Indeed its infrared luminosity $L_{IR} \sim 10^{12}\ L_\odot$ places it on the boundary between LIRGs and their bigger brothers ULIRGs (Ultra-Luminous Infrared Galaxies). Physical conditions pertaining to the starburst activity (both gas and stars) have been under active investigation via infrared spectroscopy (e.g. Tecza et al. 2000; Ohyama et al. 2000; Ohyama et al. 2003; Lutz et al. 2003) as well as via radio and mm-wave interferometry (e.g. Beswick et al. 2001; Gallimore & Beswick 2004; Tacconi et al. 1999). A superwind is seen in H$\alpha$ emission (Heckman Armus & Miley 1987, 1990; Armus et al. 1990) and in $H_2$ line emission (Wright et al. 1984, Joseph et al. 1984), presumably driven by the starburst activity. At infrared wavelengths, NGC 6240 has the most powerful galactic $H_2$ line emission known to date: $L(H_2) \sim 2 \times 10^9\ L_\odot$ (Joseph et al. 1984).

Because of the relative proximity of NGC 6240, the physical relationship between its stellar populations and gas can now be studied at high spatial resolution. One second of arc corresponds to 470 pc, so that scales of ~ 25 pc have been probed using visible-light (Rafanelli et al. 1997; Gerssen et al. 2001 and 2004) observations with the Hubble Space Telescope. Scales of ~ 50 pc are probed using adaptive optics on 8-10 m telescopes and using the NICMOS camera on HST, as reported here and by Scoville et al. 2000. Near infrared continuum emission in the nuclear regions arises from a combination of stellar continuum and warm dust. $H_2$ line emission arises when molecular hydrogen gas is excited by shocks or by hard photons; both excitation sources can be produced, in principle, either by the starburst or by the AGNs.

In this paper we use high spatial resolution observations in the near infrared from both Keck Telescope adaptive optics and NICMOS with the Hubble Space Telescope to address the following pieces of the puzzle of how the two AGNs, the merger event, and the starburst activity are related:

1) Geometry, extinction, and relative positions of the two nuclei: What is the infrared morphology of the two nuclei? What is the relation between nuclear features seen in the near infrared and those in the optical? Which features in the near-infrared and optical correspond to the radio and hard x-ray point sources?

2) Geometry for the excited gas and winds: What is the relationship between the H$\alpha$-emitting gas and the H$_2$-emitting gas? Does the excited molecular hydrogen seen in the near-infrared relate to the dense concentration of CO observed by Tacconi et al. (1999) between the two nuclei? How do the observed regions of excited H$_2$ and H$\alpha$-emitting gas fit in to the scenario in which "bubbles" of hot gas seen in low-energy x-rays (Komossa et al. 2003) occupy the regions surrounding the double nucleus?

2. TELESCOPES AND INSTRUMENTS

2.1 Keck Telescope

The Keck II adaptive optics system (Wizinowich et al. 2000a and 2000b, Johansson et al. 2000) is located on an optical bench at the Nasmyth platform of the 10-m Keck II telescope. The Xinetics deformable mirror has 349 degrees of freedom, of which approximately 249 are illuminated at any given time as the hexagonal pupil of the telescope rotates on the round deformable mirror. The Shack-Hartmann wavefront sensor is based on a 64x64 pixel Lincoln Laboratories CCD with read noise of approximately 6 electrons $px^{-1}$. The real-time computer is based on the Mercury RACE architecture, and uses sixteen Intel i860 floating-point CPUs.

For the observations reported here, the wavefront sensor's sample rate was 90-100 Hz on a guide star of magnitude R = 11.9, B = 13.5, located at a distance of 35.8 arc sec from the midpoint between the dual nuclei; for these parameters the control system's closed-loop bandwidth was typically 5 – 10 Hz. The measured Strehl of the guide star itself was ~ 45% when these observations were taken. We could not measure the Strehl at the position of NGC 6240's dual nuclei, because the wings of the point-spread functions of the numerous point-like sources in the nucleus are lost in the general diffuse light from the two galaxies. If we use the theory of (infinite-aperture) anisoplanatism assuming Kolmogorov turbulence, and estimate the Mauna Kea K-band isoplanatic angle as $\theta_0$ ~ 30 arc sec, then we would estimate the Strehl at the nucleus of NGC 6240, 35.8 arc sec from the guide star, as $S_{nuc}/S_{gs} = \exp-(\theta/\theta_0)^{5/3}$ or $S_{nuc}$ ~ 10%. Without the benefit of a real-time PSF estimator for Shack-Hartmann sensing that is analogous to that developed by Véran for curvature sensing (Véran et al. 1997) we do not have a direct measurement of the off-axis Strehl ratio.

The NIRC-2 instrument that we used at Keck (Principal Investigator: K. Matthews) is a near infrared camera providing high-resolution imaging in conjunction with the adaptive optics system on the Keck II telescope. The instrument features a Boeing 1024 x 1024 Aladdin-3 InSb detector array. Three cameras are available, providing fields-of-view measuring 10, 20, and 40 arc sec on a side, with nominal plate scales of 0.01, 0.02, and 0.04 arc sec $px^{-1}$ respectively.

These correspond to Nyquist-sampling the diffraction-limited full width at half maximum for imaging at 1, 2, and 4 $\mu$m. For the work reported here we used the "wide" camera (40 arc sec field of view, 0.039686 arc sec $px^{-1}$). NIRC2 also provides diffraction-limited grism spectroscopy, which we shall report upon elsewhere.

2.2 Hubble Space Telescope

On HST, the NICMOS camera employs NICMOS-3/PACE 256x256 HgCdTe detectors, which in Camera 2 (its "intermediate" field camera, 19.2 arc sec x 19.3 arc sec) samples the f/45 reimaged 2.4-meter primary mirror focal plane at a scale of 0.076 arc sec $px^{-1}$. See Thompson et al. 1998 and Roye et al. 2003 for details of the NICMOS instrument, and its performance on HST. The spacecraft can be repointed with high precision, enabling very fine sub-pixel dithers (to <0.1 pixel precision in Camera 2) to better sample the HST + instrumental point spread function.

3. OBSERVATIONS

Table 1 summarizes our Keck adaptive optics imaging observations. A natural guide star was used as the wavefront reference source. This star, coordinates $16^h53^m01".218$ +02 $24^m14".49$ (J2000.0) has magnitude R = 11.9, B = 13.5, and is 35.8" away from the mid-point between the two nuclei of NGC 6240. With NIRC2, we observed the NGC 6240's double nucleus in three broadband filters: K' (central wavelength $\lambda_c$ = 2.124 $\mu$m, FWHM $\Delta\lambda$ = 0.351 $\mu$m), H ($\lambda_c$ = 1.633 $\mu$m, $\Delta\lambda$ = 0.2146 $\mu$m), J ($\lambda_c$ = 1.248 $\mu$m, $\Delta\lambda$ = 0.163 $\mu$m); and two narrowband filters: nominal Br$\gamma$ ($\lambda_c$ = 2.1642 $\mu$m, $\Delta\lambda$ = 0.0339 $\mu$m), and nominal $H_2$ (called

NB2108 on the Keck web pages; $\lambda_c$ = 2.1244 $\mu$m, $\Delta\lambda$ = 0.0321 $\mu$m). The filter transmission functions for the Keck Br$\gamma$ and H$_2$ filters are shown in Figure 1a. These two filters were chosen because the strongest observed lines of H$_2$ seen in our spectra of the two nuclei and the region between them fall within the bandpass of the nominal Br$\gamma$ filter, whereas the adjacent short-wavelength continuum falls within the bandpass of the nominal H$_2$ filter.

Table 2 summarizes the raw NICMOS data that we acquired from the Hubble Space Telescope archive. The NICMOS images reported upon in this paper were obtained by recalibrating and reprocessing these archival data. Wideband F110W ($\lambda_c$ = 1.100 $\mu$m, FWHM = 0.592 $\mu$m; roughly analogous to J band), F160W ($\lambda_c$ = 1.594 $\mu$m, FWHM = 0.403 $\mu$m; similar to H-band, F222M ($\lambda_c$ = 2.216 $\mu$m, FWHM = 0.143 $\mu$m; similar to K'-band) images of NGC 6240 were obtained as part of GTO program 7219 (Scoville, PI) using the NIC2 camera, on 2 Feb 1998 in a single HST orbit. NICMOS narrowband images in the 1 – 0 S(1) transition of molecular hydrogen used the NIC2 camera with the F216N ($\lambda_c$ = 2.149 $\mu$m, FWHM = 0.020 $\mu$m) and F215N ($\lambda_c$ = 2.164 $\mu$m, FWHM = 0.021 $\mu$m) filters, GO program 7882 (van der Werf, PI) acquired 03-04 March 1998. The NICMOS F215N and F216N filters are narrower than their Br$\gamma$ and H$_2$ narrowband counterparts at Keck (compare Figures 1a and 1b). But for the present purposes this makes no difference, because our spectra (to be described elsewhere) show that emission lines in the nuclei of NGC 6240 are confined to the NICMOS F216N and Keck Br$\gamma$ filters, as shown by the red lines in Figures 1a and 1b.

NICMOS GTO program 7219 also observed NGC 6240 in a pair of narrowband filters usually used for Paschen $\alpha$ (rest wavelength 1.875 $\mu$m) whose bandpasses are shown in Figure 1c. At the redshift of NGC 6240 (z = 0.0243) the Paschen $\alpha$ line is shifted almost out of the F190N filter, as shown by the green line in Figure 1c. The 1 – 0 S(5) transition of molecular hydrogen (rest wavelength 1.8358 $\mu$m) is redshifted squarely into the F187N filter. So the difference between the F187N (emission line) and F190N (continuum) filters gives an image of the core of NGC 6240 in the $H_2$ 1 – 0 S(5) transition. However because the Pa$\alpha$ line can be quite strong in galaxies such as NGC 6240, we cannot completely discount its potential contribution to the difference between the fluxes in F187N and in 190N. Although the flux difference is dominated by the 1 – 0 S(5) line of $H_2$, we have chosen not to do relative photometry between this narrowband filter pair and the F215N – F216N pair on account of the modest but unknown contribution of Pa$\alpha$.

4. DATA ANALYSIS TECHNIQUES

The NIRC2 data were reduced using standard techniques for near-infrared imaging. Bad pixels were masked out when combining the dithered images to obtain a final image. Flat-fields were constructed from twilight exposures at each filter bandpass. Dark frames were created for each exposure time. Each dithered image was sky-subtracted, dark-subtracted, and flat-fielded. Deconvolution was not used for the images discussed in this paper.

NICMOS broadband images were reduced as follows. All data were taken in STEP8 multiaccum sequences: F110W with NSAMP=10 (40 sec integrations), F160W with NSAMP =11 (48 sec integrations), and F222M with NSAMP=12 (56 sec integrations). Four images were taken in each broadband filter under a 4-point square dither pattern, with 1.9125 arc sec offsets on each side of the square. Offsets by non-integral pixel sizes allow for better sampling of the PSF, and the four images in each filter were median combined on a 2x re-sampled grid (0.038 arc sec px$^{-1}$) after image re-registration. During the image registration/combination process, a +0.9% linear geometrical distortion (image Y:X) in NICMOS Camera 2 was corrected. The total target flux density was preserved in the re-sampling so that the flux density pixel$^{-1}$ area is conserved.

The narrowband F215N and F216N images obtained as part of the van Der Welf GO program were not dithered. The narrowband images were taken with MIF512/NSAMP=25 multiaccum sequences (integration time 512 seconds). Eight images were taken in the F216N filter (4096 sec total integration time), six images were taken in the F215N filter (3072 sec total integration time). After each exposure the telescope was chopped off the source by 90 arc sec and sky background frames were taken in an identical manner. After recalibrating the raw frames (with post-processing to interpolate and replace known bad pixels) to create count rate images, the sets of filtered target and background images were median combined.

The NICMOS narrowband F187N and F190N images were taken in GTO 7219 in the same spacecraft orbit as the broadband images (above), and under the same 4-point dither pattern. At each dither point (for each filter) a single STEP8/NSAMP=12 exposure (56 sec integration) was taken, yielding 218 sec total integration time after combining the four images. The data were processed in a manner similar to the broadband images. Measurements taking the pixel sampling into account yield a NICMOS Strehl ratio of ~ 98% in the F110W filter and the NIC2 camera used here.

## 5. RESULTS AND INTERPRETATION
### 5.1 Broadband Imaging

Figure 2a shows a false-color wide-field image of NGC 6240 and its guide star (on the left), from Keck adaptive optics. This image is 59 arc sec wide by 47 arc sec tall. Red corresponds to K'-band, green to H-band, and blue to J-band. One can clearly see light from the extended parts of the galaxy in J-band (blue), with the characteristic "bow-tie" shape known from previous work (e.g. Figure 2 in Pasquali et al. 2003). The compact nuclear region is dominated by K'-band light (red in our Figure 2); this is consistent with previous estimates of high dust extinction in the double nucleus (reviewed by Gerssen et al. 2004). Note that the guide star (left) is overexposed in this image, and that it shows the six-fold symmetry expected from the Keck hexagonal pupil.

Figures 2b and 2c show false-color images of NGC 6240 over a slightly narrower field, 28.5 x 33 arc sec. Here red represents the Keck K' adaptive optics image, green is F814W WFPC2 archival data from HST, and blue is F450W archival WFPC2 data. In these images spanning the wavelength range 450 nm to 2.1 microns, the highly structured dust lanes cutting through or in front of the central regions of NGC 6240 are clearly seen. Figure 2b has a linear color map; 2c has a logarithmic color map so that the position of the double nucleus can be identified.

Figure 3 zooms in still further, showing the double nucleus of NGC 6240 in J, H, and K' bands from Keck adaptive optics (frames a, b, c) and in F110W, F160W, and F222M from NICMOS (frames d, e, f). North is up and East is to the left. The Keck images span 4.2 arc sec in the North-South direction, corresponding to 1.9 kpc at the distance of NGC 6240; the NICMOS images are 3.9 arc sec in the North-South direction, corresponding to 1.8 kpc.

Before we discuss the astrophysical content of these images, it is instructive to compare the technical performance of ground-based adaptive optics (AO) with that of NICMOS on HST. First and foremost, we note that all the major features of the two nuclei are seen in both Keck AO and NICMOS. This is distinctly reassuring, in view of the potential for AO images to contain artifacts due to imperfect calibration of non-common-path errors.

In terms of the more detailed sub-structures within each nucleus and between the two nuclei, the Keck AO and NICMOS images contain very complementary information. Close examination of the data at all three wavelengths indicates that the two main nuclei are surrounded by a multitude of unresolved point-like sources. One can see from Figure 3 that these point-like sources have highest contrast in the Keck K' image (c) and in the NICMOS F110W (~J) image (d).

This qualitative impression from Figure 3 can be made more quantitative. Figure 4 shows images of one of the point-like sources, located at $16^h52^m58".851$, $+02\ 24^m09".75$ (J2000.0), at three wavelengths with both Keck and NICMOS. The top row shows the Keck adaptive optics images with (a) J filter, (b) H filter, and (c) K' filter. The bottom row shows the NICMOS images with (d) the F110W, (e) F160W, and (f) F222M filters. In each image, the dynamic range between black and white is a factor of 10 (actually 10 ± 1, due to the vagaries of image reproduction). The compactness of the (black) core relative to the ten-times lower (white) background is a measure of the imaging performance. The numbers listed on Figure 4 and in Table 3 are the full width at half maximum (FWHM) of the respective point-source images.

One sees from Figure 4 that for NICMOS, the FWHM improves at shorter wavelengths, whereas for Keck AO the reverse is the case. This is because the imaging performance of a diffraction-limited telescope such as HST improves as the wavelength gets <u>shorter</u>, whereas the Strehl ratio for a given adaptive optics system improves as the wavelength gets <u>longer</u>. Both of these effects are seen in Figure 4, in which NICMOS has superior spatial resolution at F110W (J-band) and Keck has better spatial resolution at K-band.

In principle one would expect a diffraction-limited Keck telescope, at 10 m diameter, to show more than 4 times better spatial resolution than NICMOS on HST. This increased spatial resolution is indeed achievable when 1) the guide star is bright (e.g. V < 10), 2) the guide star is close by (e.g. < 10 arc sec), and 3) the pixel sampling used for the NIRC2 camera is appropriate for Nyquist-sampling the diffraction limit at each wavelength. However for the present observations none of these conditions was satisfied: the guide star was >13th magnitude, it was ~35 arc sec away, and the 0.04-arc sec pixels on NIRC2's wide camera do not Nyquist-sample any of the J, H, K wavelengths reported here.

A quantitative measure of point-source sensitivity comes from analyzing the signal-to-background ratio for the point-like source shown in Figure 4, as well as for a second point-source at $16^h52^m58".684, +02\ 23^m56".9$, J2000.0 (point source 2 in Table 3). Table 4 shows the signal-to-background ratios calculated for these two point sources for all three wavelengths bands of Keck AO and NICMOS data. For both sources NICMOS has superior signal-to-background ratio at J and H bands, whereas Keck AO has superior signal-to-background ratio at K band/F222M.

We now return to the features seen in the two nuclei of NGC 6240 (Figure 3). The most striking impression is that both the North and South nuclei are elongated, with considerable sub-structure within each nucleus. The brightest point in the North nucleus is to the northeast. As shown in Figure 5, we shall refer to this as North1. The brightest point in the South nucleus is to the north-northwest; we shall refer to this as South1. Just outside the two nuclei are many unresolved point-like sources that we propose are young star-clusters, participating in the starburst in a manner similar to the clusters in the extra-nuclear regions of the NGC 6240 studied by Pasquali et al. (2003). Authors such as Zepf et al. (1999), Schweizer and Seitzer (1998), and Whitmore & Schweizer (1995) have shown in other spiral-galaxy mergers that point-like sources analogous to these "have sizes, colors, and luminosities like those expected for young Galactic globular clusters." Figure 6 shows Keck AO J, H, and K' images of the nuclei of NGC 6240, with grey scale stretched so as to emphasize the point-like sources just outside the two nuclei. The nature of the point-like putative clusters in our images will be discussed in a future paper.

There are clear wavelength-dependent differences in morphology within both nuclei. First consider the South nucleus in Figure 3. In Keck J-band (Frame a) and NICMOS F110W (Frame d) there are two clear intensity peaks within this nucleus, as there are at 814 nm (Gerssen et al. 2004, discussed further below). But at longer wavelengths (H and K') the primary peak, South1, becomes brighter relative to the secondary peak, which we shall call South2 (see Figure 5). This can be seen more clearly in Figure 7 panels (g) and (h). From the Keck data using photometry with a 3-pixel-radius aperture, the ratio of flux in South1 to flux in South2 is 1.23 at J-band, 1.54 at H-band, and 2.07 at K'-band.

In the North nucleus, Keck adaptive optics data in K' band show a new point-like source emerging to the southwest of North1 (Figure 7d and Figure 5). This second point source is suggested in the NICMOS F222W image (Figure 3f), but is more clearly differentiated in the Keck K' data. From the Keck data using photometry with a 3-pixel-radius aperture, the ratio of flux in North2 to flux in North1 is 0.57 at J-band, 0.60 at H-band, and 0.68 at K'-band. (In the J and H band photometry, we constrained the "North2" aperture to be centered at the centroid of North2 at K'-band.)

A three-color image synthesizing these data on the nuclear region is shown in Figure 8, which uses a square-root color map. The NICMOS F110W image is shown in blue; the Keck H and K' images are in green and red respectively. Light from the extended regions of the merging galaxies appears bluish in this image. Areas of high dust extinction appear selectively in red in Figure 8, and match those seen in Figures 2b and 2c.

The region of highest reddening extends north of the South1 nucleus, reaching into the gap between the North and South nuclei. In the CO map of Tacconi et al. (1999, Figure 1), the South1 nucleus lies at the southern edge of a region of high-density CO. Analogously, in our Keck data the South1 nucleus lies at the southern edge of a highly reddened region that we associate with Tacconi's CO cloud.

The most heavily reddened feature is the South1 nucleus, suggesting that it is the most deeply embedded in dust. We also note an "arm" of high reddening that wraps around the South nucleus from northeast to southwest, separating the South1 and South2 sub-nuclei. This "arm" within the South nuclear region has the appearance of a foreground dust lane, obscuring the bright regions behind it.

5.2 Differential extinction toward the two nuclei, and the location of the two AGNs

Many authors have pointed out that the angular separation between the North and South nuclei is smaller at infrared wavelengths than in the visible. It is smaller still at mm and radio wavelengths. The angular separation measured by Gallimore and Beswick (2004) between the two brightest components was 1.52 arc sec in their 5 GHz MERLIN radio data. The radio continuum properties of the two brightest nuclei seen in the 5 GHz MERLIN data resemble those of compact radio sources in Seyfert nuclei (including an inverted spectrum at low frequencies, and high brightness temperatures). The distance between the two brightest MERLIN components is hence a good measure of the actual distance between the two black holes. Gerssen et al. (2004) measured the angular separation between the North and South nuclei as 1.86 arc sec at V band, and suggested that non-uniform dust obscuration of the South nucleus, and possibly the North nucleus as well, might explain the wavelength-dependence of the angular separation.

We have shown our Keck AO data in J and K' band in Figure 7, along with archived WFPC2 data obtained by Gerssen et al. in the F450W filter (the so-called "wide B" filter, centered roughly where the B band is centered, but about twice as wide) and F814W filter (roughly Johnson I band). In the South nucleus one sees the northwest component, which we have called South1 and which Gerssen et al. called N2, becoming more prominent relative to the southeast component (our South2 and Gerssen et al.'s N1) at longer wavelengths. The effect is quite striking: at F450W South1 is barely visible, whereas at K'-band South1 is by far the more prominent sub-component. This is consistent with the heavy reddening of South1 shown in Figure 8. In the North nucleus, the southwest component North2 is emerging as a point-like source in K'-band, whereas at F450W it is not visible at all.

There are two effects that make the angular separation between the North and South nuclei appear to be smaller at K'-band than at F450W. At K' band, the photocentric location (i.e. the image centroid) of the Southern nucleus is dominated by the contribution of the South1 source. At the same time, the emergence of North2 in K'-band biases the photocentric location of the North nucleus farther to the south than at shorter wavelengths. The measured separation between the centroids of the North1 and South1 sub-nuclei in the HST F450W filter is 1.87 arc sec. At K' band, where we begin to see the North2 sub-nucleus clearly, the separation between the North2 and South1 sub-components is considerably smaller: 1.59 ±0.02 arc sec.

We note that the K'-band angular separation of the two nuclei, though smaller than that at shorter wavelengths, is not as small as the measured separation of 1.52 arc sec at 5 GHz (Gallimore and Beswick, 2004). Thus we conclude that the North2 and South1 nuclei are much more likely to be close to the two AGNs than the North1 and South 2 sub-nuclei that are so prominent in the F450W filter. However because our 1.59 arc sec K'-band separation between N2 and S1 is still not as small as the 1.52 arc sec separation seen at 5 GHz, we agree with Gerssen et al. that even at K'-band we are not completely penetrating the large amount of dust extinction: either one or both AGNs have not fully emerged from the dust in our K'-band images.

If the unresolved point-source at North2 is the site of one of the black holes in NGC 6240, then the second black hole would lie north of South1 by about 0.07 arc sec, or approximately two pixels in the NIRC2 wide camera, based on the angular separation of 1.52 arc sec between the two nonthermal radio sources at 5 GHz measured by Beswick .

5.3 Narrow-band imaging in lines of molecular hydrogen

It has been known for more than 15 years that NGC 6240 has extended filamentary Hα emission over a region up to 50 kpc in size, which Heckman et al. (1987, 1990) interpreted as being due to a superwind driven by vigorous starburst activity. A recent paper by Veilleux et al. (2003) reinforces the interpretation that [NII] and H$\alpha$ optical line emission from excited gas in NGC 6240 is associated with a superwind, presumably driven by the merger-induced starburst. NGC 6240 is also an extremely energetic emitter of near-infrared molecular hydrogen lines. With the high spatial resolution available via adaptive optics and NICMOS, one can use infrared narrow-band imaging to gain insight into the spatial distribution of excited $H_2$ gas, to compare this with the spatial distributions of H$\alpha$ and of soft x-rays, and to integrate observations at these disparate wavelength bands into a more unified understanding of gas emission in starburst galaxies.

The physical processes that contribute to the complex spatial structure of molecular gas in NGC 6240 have been emerging over the past few years from observations with spatial resolution 0.5 – 1 arc sec. Tacconi et al. (1999) reported observations with the IRAM mm-wave interferometer of the 1.3 mm CO J = 2 $\rightarrow$ 1 line in the core of NGC 6240, at a spatial resolution of 0.5 x 0.9 arc sec. Tacconi et al. interpreted their data as suggesting a rotating, turbulent thick molecular disk structure about 500 pc in diameter lying between the two nuclei (centered about 0.6 arc sec north-northeast of the South nucleus). Ohyama et al. (2000) used long-slit K-band spectroscopy with the Subaru telescope in seeing-limited mode, and found a peak in $H_2$ 1-0 S(1) line emission between the two nuclei, but a bit closer to the South nucleus (0.4 arc sec north-northeast of the South nucleus) than Tacconi's reported CO peak. In addition to the CO and $H_2$ structures between the two nuclei, a series of narrow-band images by Sugai et al. (1997) and Herbst et al. (1990) under seeing-limited conditions, and by Bogdanovic et al. (2003) with adaptive optics, showed that there are diffuse extensions of the $H_2$ emission in directions southwest and southeast of the South nucleus.

Because of the fortuitous redshift of NGC 6240 ($z = 0.0243$), the $H_2$ 1-0 S(1) emission line falls within the bandwidth of most standard narrowband Brγ filters. By imaging the $H_2$ 1-0 S(1) emission line in the nominal narrowband Brγ filter, and subtracting the adjacent continuum in a narrowband filter centered at 2.124 μm, one can obtain images of NGC 6240 largely in light from the strong $H_2$ 1-0 S(1) emission line. Bogdanovic et al. (2003) used this type of narrow-band imaging with the Lick Observatory adaptive optics system to delineate the structure of the extended $H_2$ emission regions south-west and south-east of the South nucleus. We report here on analogous observations at higher spatial resolution and higher signal-to-noise ratio.

Figure 9 shows the continuum-subtracted $H_2$ images from a) Keck adaptive optics, and b) NICMOS on HST. "X" marks the position of the brightest pixel in broadband images of each of the two nuclei. Striking features of the $H_2$ emission include the following:

- <u>A "ribbon" of excited molecular hydrogen</u> extending between the North and South nuclei. Figure 10 shows a zoomed-in section of Figure 9a highlighting the region between the two nuclei. The $H_2$ "ribbon" between the nuclei has a reverse S-shape. It begins in a point-source coincident with the South nucleus (discussed below), departs the South nucleus toward the east, then curves in a long loop toward the north-west before turning east again to the North nucleus. The Keck data show a hint that the "ribbon" may be subdivided into three separate "blobs". The $H_2$ "ribbon" occupies roughly the same region of projected sky as the peak of CO J = 2 → 1 line emission noted by Tacconi et al. (1999) and by Ohyama et al. (2003). But for $H_2$, at the higher spatial resolution of Keck adaptive optics and NICMOS, the inter-nuclear emission is oriented in along a diagonal northwest-southeast line, in contrast to the north-south orientation of CO emission reported by Tacconi et al. at lower spatial resolution. Figure 11 shows the narrowband image from NICMOS in the $H_2$ 1 – 0 S(5) line (the difference between images in the F187N and F190N filters of NICMOS). Although the 1 – 0 S(5) line is at lower signal-to-noise ratio than the 1 – 0 S(1) line shown in Figures 9 and 10, the morphology of these two molecular hydrogen emission-line images is quite similar; the 1 – 0 S(5) line also shows a "ribbon-like" structure between the two nuclei.

The relation between the $H_2$ "ribbon" seen by Keck and NICMOS and the disk of dense CO and neutral hydrogen hypothesized by Tacconi et al. (1999) and by Beswick et al. (2001) is not clear. Perhaps the diagonal "ribbon" is formed at the locations where flowing or expanding molecular hydrogen gas from the South nucleus crashes into the outer layers of Tacconi et al.'s North-South oriented cloud of CO lying between the two nuclei of NGC 6240.

However an alternative explanation is also possible. Computer simulations (e.g. Barnes and Hernquist 1991, 1996) of disk galaxy mergers show that "a significant amount of material flows along the bridge connecting the galaxies..." (Barnes & Hernquist 1991; see also Figures 9 and 10 in Barnes and Hernquist 1996, and Figures 1 and 4 in Barnes 2002). The morphology of the $H_2$ "ribbon" seen by Keck and NICMOS suggests that we may indeed be seeing gas flowing from one nucleus to the other, with an elongated and irregular shape determined both by the initial parameters of the galaxy merger and by tidal distortion of the "bridge" between the two galaxies. Future spatially resolved adaptive optics spectroscopy will help to assess these varying hypotheses.

- <u>Compact knots of $H_2$ emission</u> to the northeast of the North nucleus, to the east-southeast of the South nucleus, and to the south-southwest of the South nucleus. Each of these compact knots corresponds to one of the CO features noted by Tacconi et al. (1999) as D, C, and A in their Figure 1, and were noted by Bogdanovic et al. (2003) in Lick adaptive optics data. These features were shown by Tacconi et al. to have distinct CO velocity structure, and hence they are presumably distinct dynamical entities. A plausible hypothesis is that these are star-forming or starburst regions, a hypothesis that can be tested via future adaptive optics spectroscopy of each knot.

- <u>Point-like $H_2$ emission at the South nucleus.</u> Figure 9 shows that in both the Keck AO

and NICMOS images we see point-like $H_2$ emission at the position of the South nucleus (South1). We measured the centroid of the bright point corresponding to the South1 sub-nucleus in the continuum-subtracted $H_2$ image, and compared it with the centroid of South1 in the continuum image. For the Keck adaptive optics data, these positions were identical to within less than a pixel (<0.04 arc sec), consistent with the measurement error. These data are consistent with the hypothesis that the positions of the peak $H_2$ emission and of the peak continuum emission are coincident to within the diffraction limit of the telescope. We did the same exercise for the $H_2$ emission from the North1 and North2 nuclei, evident in Figure 10. As for the South1 sub-nucleus, the peak $H_2$ emission from North1 and North2 corresponded in position to that in the continuum image.

Using seeing-limited observations, previous authors (Tecza et al. 2000, Herbst et al. 1990, van der Werf at al. 1993, Sugai et al. 1997, Ohyama et al. 2000) have reported that the peak of the $H_2$ is displaced 0.35 – 0.4 arc sec north of the South nucleus, in the direction of the North nucleus. Our high-resolution data from both Keck adaptive optics and NICMOS are not consistent with this displacement. We point out that the complex ribbon-like spatial structure of the $H_2$ emission seen in Figures 9 and 10 might cause a seeing-limited image to have a centroid that is displaced northward of the South nucleus.

- <u>Filamentary structure in molecular hydrogen.</u> In the region <u>exterior</u> to the two nuclei of NGC 6240, the filaments of excited $H_2$ observed by Keck adaptive optics and by NICMOS correspond well to the filaments observed in $H\alpha$ emission by WFPC2 on the Hubble Space Telescope. This can be seen in the contour plots of $H_2$ and $H\alpha$ shown in Figures 12a and 12b respectively. Referring to the CHANDRA observations of Komossa et al. (2003), shown as a contour plot in Figure 13, the two largest $H_2$ and $H\alpha$ filamentary loops in Figure 12 to the east and west of the North nucleus trace the exterior boundaries of two "bubbles" of soft x-ray emission observed by CHANDRA. The $H\alpha$ emission around the circumference of these bubbles (Fig. 12b) persists to larger distance from the center of the nuclear region than does the $H_2$ emission (Fig. 12a), possibly indicating that the molecular clouds are more concentrated in the nuclear region. Komossa et al. (2003) suggest that the extended soft x-ray "bubble" features are filled with hot (kT = 0.8 – 2.8 keV) low-density (0.1 particle/cm$^3$) gas and are powered by starburst-driven superwinds with mechanical input power from ~3 supernovae per year over a starburst duration of 3 x 10$^7$ years. In this scenario the $H_2$ and $H\alpha$ line emission shown in Figure 12 would come from a thin layer around the edges of these extended soft x-ray bubbles, in locations where the starburst wind is driving shocks or ionization fronts into the interstellar medium and surrounding molecular clouds.

6. CONCLUSIONS

We have presented results of near infrared imaging of NGC 6240 using adaptive optics on the Keck II Telescope and NICMOS on the Hubble Space Telescope. Broadband Keck images were obtained in J, H, and K' bands. NICMOS images were in filters approximately corresponding to those at Keck. In addition, narrowband images were obtained in the 1-0 S(1) and the 1 – 0 S(5) emission lines of molecular hydrogen.

In general we find very good agreement between the Keck AO and NICMOS data. The Keck AO images have somewhat higher spatial resolution in K band, whereas NICMOS is superior in F110W (roughly analogous to J band). On this basis we foresee continued strong synergy between ground-based adaptive optics on 8 – 10 m telescopes, and space-based near infrared observations on somewhat smaller telescopes.

Both the North and South nuclei of NGC 6240 are clearly elongated, with considerable sub-structure within each nucleus. In K' band there are at least two point-sources within the North nucleus; we tentatively identify the south-western point-source as the position of one of the two AGNs. In the South nucleus, the northern sub-nucleus is the most highly reddened. Using the nuclear separation measured in the radio at 5 GHz we suggest that the AGN in the South nucleus is still enshrouded in dust at K' band, and is located approximately 0.07 arc sec to the north of the K' band position of the northern subnucleus.

<u>Within</u> the South nucleus there is strong point-like $H_2$ 1-0 S(1) line emission from the northern subnucleus, contrary to the conclusions of previous seeing-limited observations. Narrowband $H_2$ emission-line images show that a streamer or ribbon of excited molecular hydrogen connects the North and South nuclei, on a diagonal from southeast to northwest. We suggest that this linear feature corresponds to a bridge of gas connecting the two nuclei (and perhaps flowing between them), as seen in computer simulations of mergers. The connection between this streamer and the "disk" of CO seen at lower spatial resolution by previous observers is not yet clear.

Finally, many point-like regions are seen around the two nuclei. These are most prominent in J-band with NICMOS, and in K'-band with Keck adaptive optics. We suggest that these point-sources represent star clusters formed in the course of the merger. In future observations it will be of great interest to determine the ages of these star clusters, and to determine what fraction of NGC 6240's starburst activity can be attributed to them.

ACKNOWLEDGEMENTS

We enthusiastically thank the staff of the W. M. Keck Observatory, and especially David Le Mignant and the other members of its adaptive optics team, for their dedication and hard work. Data presented herein were obtained at the W. M. Keck Observatory, which is operated as a scientific partnership among the California Institute of Technology, the University of California, and the National Aeronautics and Space Administration. The Observatory and the Keck II adaptive optics system were both made possible by the generous financial support of the W.M. Keck Foundation. The authors wish to extend special thanks to those of Hawaiian ancestry on whose sacred mountain we are privileged to be guests. Without their generous hospitality, the observations presented herein would not have been possible.

This work was supported in part under the auspices of the U.S. Department of Energy, National Nuclear Security Administration by the University of California, Lawrence Livermore National Laboratory under contract No. W-7405-Eng-48. This work was supported in part by the National Science Foundation Science and Technology Center for Adaptive Optics, managed by the University of California at Santa Cruz under cooperative agreement No. AST-9876783. D. W. and R. A. acknowledge a mini-grant from the Institute of Geophysics and Planetary Physics at the Lawrence Livermore National Laboratory. The work by R.A. was supported in part by NSF grant AST-0098719. The NICMOS observations reported herein were obtained with the NASA/ESA Hubble Space Telescope operated by the Space Telescope Science Institute managed by the Association of Universities for Research in Astronomy Inc. under NASA contract NAS5-26555. This work was also supported in part by NASA grants NAG 5-3042 and NAS 2-26555 to the NICMOS IDT.

FIGURE CAPTIONS

Figure 1. Narrowband filters. (a) Bandpass plots of Brγ and $H_2$ (NB2108) filters for the NIRC2 adaptive optics camera at the Keck II telescope. The $H_2$ 1-0 S(1) line of NGC 6240 falls entirely within the Brγ filter, as shown by the red horizontal bar. (b) Bandpass plots of NICMOS F215N and F216N filters. The $H_2$ 1-0 S(1) line of NGC 6240 falls entirely within the F216N filter. (c) Bandpass plots of NICMOS F187N and F190N filters. The $H_2$ 1-0 S(5) line of NGC 6240 falls largely within the high-sensitivity portion of the F187N filter. The Paschen α line is redshifted largely, but not entirely, out of the F190N filter. In these figures the width of the red horizontal bars ($H_2$ lines) were estimated from spectra (reported elsewhere). The width of the Paα line (Figure 1c) was estimated from that of the Brγ line seen in our spectra.

Figure 2. Multi-wavelength images of NGC 6240. North is up and East is to the left in all figures in this paper. (a) False-color wide-field image of NGC 6240, with Keck adaptive optics. K'-band is represented by red, H-band by green, and J-band by blue. The (over-exposed) guide star is on the far left of this image, and is about 36 arc sec from the double nucleus. This image is 59 arc sec wide by 47 arc sec tall. The large-scale shape of the galaxy as seen in J-band (blue) light is similar to the bow-tie shape long known in the visible (see Figure 2 in Pasquali et al. 2003). (b) False-color image of NGC 6240 over a narrower field (28.5 x 33 arc sec) than in Figure 2(a) as seen by WFPC2 PC chip on HST and with Keck adaptive optics. This is a linear color map; red represents K'-band from Keck adaptive optics, green is the F814W filter from WFPC2 on HST, and blue is the F450W filter from WFPC2 on HST. The large dust lanes in (or in front of) the nuclear area are clearly seen. c) Same as b) but a logarithmic color map so that the double nuclei are not saturated.

Figure 3. The double nucleus of NGC 6240. North is up and East is to the left. Top three panels show Keck adaptive optics images in (a) J, (b) H, and (c) K' bands. Bottom three images show NICMOS images to approximately the same spatial scale: (d) F110W filter, (e) F160W filter, (f) F222M filter. Keck images are ~ 4.2 arc sec in the north-south (up-down) direction. NICMOS images shown here are ~3.9 arc sec in the north-south direction. The color map is logarithmic. Each image has been re-scaled so that the brightest pixel in its south nucleus is white, in order to preserve dynamic range in the display. The two color bars shown for concreteness on the right of this Figure correspond to panels c) and f).

Figure 4. Images of a point-like source located at $16^h52^m58".851, +02\ 24^m09".75$ (J2000.0) in three different wavelength bands. North is up and East is to the left. The top row shows the Keck adaptive optics images with (a) J filter, (b) H filter, and (c) K' filter. The bottom row shows the NICMOS images with (d) the F110W, (e) F160W, and (f) F222M filters. These images are on a log greyscale. In each image, the dynamic range between black and white is a factor of 10 (actually 10 ± 1, due to the vagaries of image reproduction). The compactness of the (black) core relative to the ten-times lower (white) background is a measure of the imaging performance. The numbers listed on Figure 4 are the full width at half maximum (FWHM) of the respective point-source images. The overall size of the Keck panels is ~1.1 arc sec wide by 0.8 arc sec tall. The NICMOS panels are ~1.0 arc sec wide by 0.8 arc sec tall.

Figure 5.  Positions of the northern and southern sub-nuclei referred to in the text from Keck K'-band data.

Figure 6. Keck AO images of the core of NGC 6240, using a greyscale stretch that overexposes the nuclei and emphasizes the numerous point-like sources in the nuclear region.  Left Panel: J-band image, Middle Panel: H-band image, Right Panel: K'-band image.  These panels are 4.2 arc sec wide by 7.4 arc sec tall.  North is up and East is to the left.

Figure 7.  Spatially dependent extinction: features within the two nuclei of NGC 6240 as a function of observing wavelength.  Upper row shows North nucleus, lower row shows South nucleus.  Panels (a) and (e) HST WFPC2, F450W filter.  Panels (b) and (f) HST WFPC2, F814W filter.  Panels (c) and (g) Keck adaptive optics, J-band.  Panels (d) and (h) Keck adaptive optics, K'-band.  Linear color map; each panel is re-scaled so that the brightest pixel is shown in white.  North is up and East is to the left.  Panels a) – d) are 1.7 arc sec wide by 1.3 arc sec tall.  Panels e) – h) are 1.7 arc sec wide by 1.5 arc sec tall.

Figure 8.  False-color F110W-H-K' image of the nuclear regions of NGC 6240, using data both from NICMOS and Keck AO.  Square root color map.  The northern part of the South1 sub-nucleus is the most heavily reddened.  North is up and East is to the left.  This image is approximately 3.8 arc sec wide by 5.2 arc sec tall.

Figure 9. Narrowband images of the nuclear region of NGC 6240, in the $H_2$ 1-0 S(1) emission line. North is up and East is to the left. Log color map to emphasize fainter features. (a) Keck adaptive optics: difference between images obtained in Brγ filter and in $H_2$ filter. White represents the peak flux, scaled to unity. Black represents fluxes < 0.011 of the peak. This image is 7.3 arc sec wide x 5.8 arc sec tall. (b) NICMOS: difference between images obtained in F216N and F215N filters. White represents the peak flux, scaled to unity. Black represents fluxes < 0.08 of the peak. This image is 7.1 arc sec wide x 5.1 arc sec tall. The two X's mark the positions of the North1 and South1 nuclei as measured in continuum images.

Figure 10. Keck adaptive optics narrowband image of the nuclear region in the $H_2$ 1-0 S(1) emission line(as in Figure 9a), zoomed in so as to emphasize structures between and immediately around the two nuclei. This image is 7.5 arc sec wide by 5.5 arc sec tall and has a log color map. The 8 contour levels are on a log scale, with the first contour at 0.2 times the maximum flux (at the South1 nucleus) and the last contour at 0.02 times the maximum flux. The black X's show the position of the brightest pixels in the Northern and Southern nuclei, as seen on the un-subtracted image in the Brγ filter. These positions correspond to the North1 and South1 sub-nuclei. The North2 sub-nucleus also is bright in the difference image.

Figure 11. Narrowband NICMOS image of the core of NGC 6240 in the $H_2$ 1 – 0 S(5) line (difference between images in F187N and F190N filters). Image uses a log scale for the color map and contour lines, and is geometrically corrected for the Y:X = 0.9% linear geometrical distortion in NICMOS Camera 2. Although the 1 – 0 S(5) line is at lower signal-to-noise ratio than the 1 – 0 S(1) line shown in Figure 9, the morphology of these two molecular hydrogen emission-line images is quite similar. This image is 8 arc sec wide by 8.2 arc sec tall.

Figure 12. Contour plots of line-emitting gas in the region outside the two nuclei of NGC 6240. Square root color map and contour scale to emphasize fainter structures. (a) $H_2$ 1-0 S(1) emission line from Keck adaptive optics; (b) H$\alpha$ emission from the WFPC2 camera on the Hubble Space Telescope (F673N filter, archival data from HST proposal number 6430, PI van der Marel). Panel a) is 20.2 arc sec wide by 16.4 arc sec tall; panel b) is 19.6 arc sec wide by 16.3 arc sec tall.

Figure 13. Contour plot of 0.5 – 8 keV x-ray emission from NGC 6240, from archived CHANDRA ACIS-S data of Komossa et al. (2003). Exposure time was 27 ksec. Square root color map and contours, to emphasize fainter features. Image is 20.2 arc sec wide by 16.7 arc sec tall.

Table 1: Keck Adaptive Optics Observations of NGC 6240

| Date | Filter | Integration time |
|---|---|---|
| 4 Aug 2002 UT | Br$\gamma$ | 1200 sec |
| 4 Aug 2002 UT | NB2108 | 1200 sec |
| 17 Aug 2003 UT | J<br>H<br>K' | 300 sec<br>270 sec<br>390 sec |

Table 2: HST NICMOS Observations of NGC 6240 Discussed Here

| Observing Program | Filter | Camera | Integration time |
|---|---|---|---|
| GTO 7219 (Scoville, PI) | F222M | NIC2 | 224 sec. |
| | F160W | NIC2 | 192 sec |
| | F110W | NIC2 | 160 sec |
| | F187N | NIC2 | 218 sec |
| | F190N | NIC2 | 218 sec |
| GO 7882 (van der Welf, PI) | F216N | NIC2 | 4096 sec |
| | F215N | NIC2 | 3072 sec |

Table 3. FWHM for Keck adaptive optics and NICMOS images, from two point-like sources

|  | "J band" pt source 1 / pt source 2 | "H band" pt source 1 / pt source 2 | "K band" pt source 1 / pt source 2 |
|---|---|---|---|
| Keck FWHM | 0.13 / 0.16 arc sec | 0.13 / 0.13 arc sec | 0.10 / 0.12 arc sec |
| NICMOS FWHM | 0.12 / 0.11 arc sec | 0.17 / 0.13 arc sec | 0.19 / 0.17 arc sec |

Note to Table 3: FWHM was measured using IDL's IMEXAM package as called in ATV. The fitting parameters were: centering box radius = 7 pixels, inner sky radius = 10 pixels, outer sky radius = 20 pixels.

Table 4.  Signal-to-background ratios for Keck AO and NICMOS images

of two point-like sources

|  | Signal to background ratios for Point Source 1: Keck AO /NICMOS | Signal to background ratios for Point Source 2: Keck AO /NICMOS |
|---|---|---|
| J band / F110W | 1.1 / 3.4 | 1.6 / 2.5 |
| H band / F160W | 1.3 / 2.4 | 1.6 / 2.3 |
| K' band / F222M | 2.6 / 0.7 | 3.6 / 0.9 |

Note to Table 4: Object counts and sky counts were measured using IDL's IMEXAM package as called in ATV.  The fitting parameters were: centering box radius = 7 pixels, inner sky radius = 10 pixels, outer sky radius = 20 pixels.  The aperture radius was chosen as that radius where the PSF has fallen to the sky background level as measured between radii of 10 and 20 pixels.  Background counts were calculated by integrating IMEXAM's sky level out to the aperture radius.

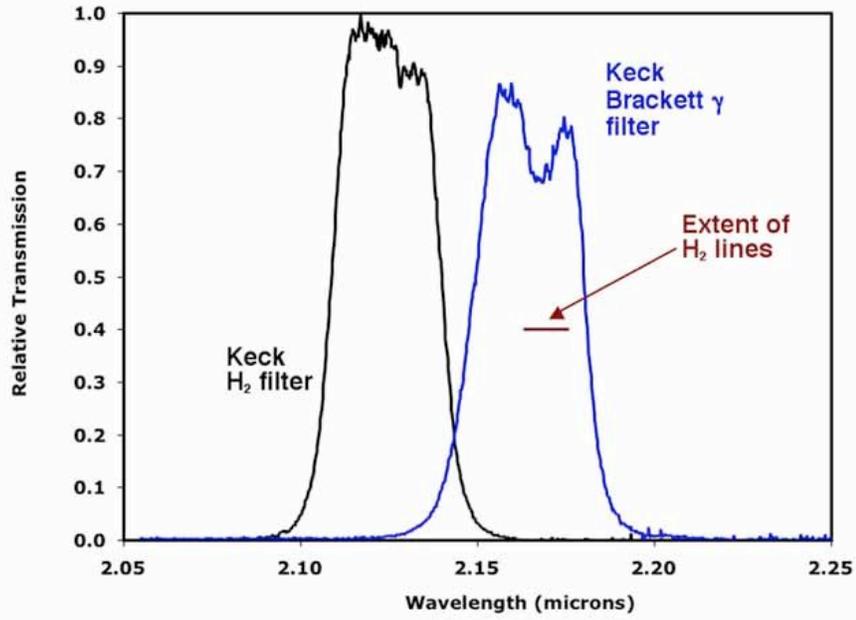

Figure 1a

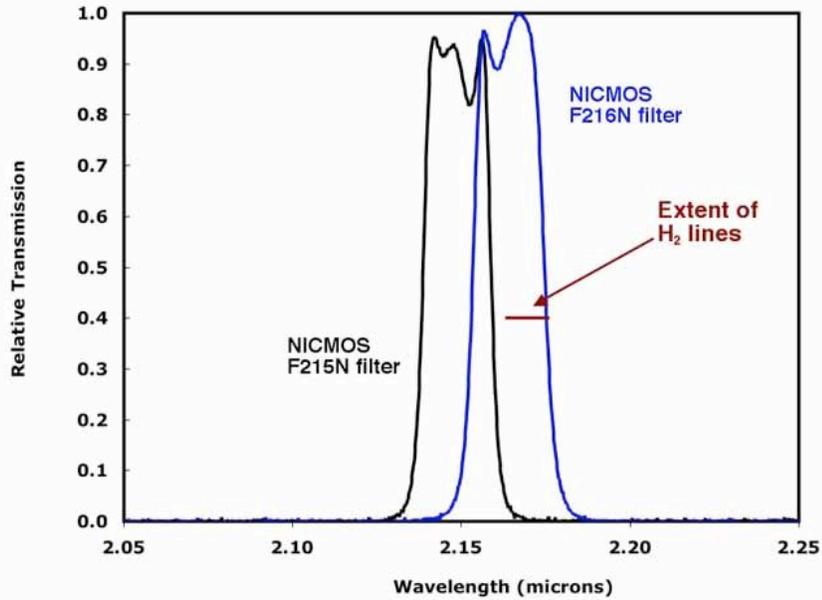

Figure 1b

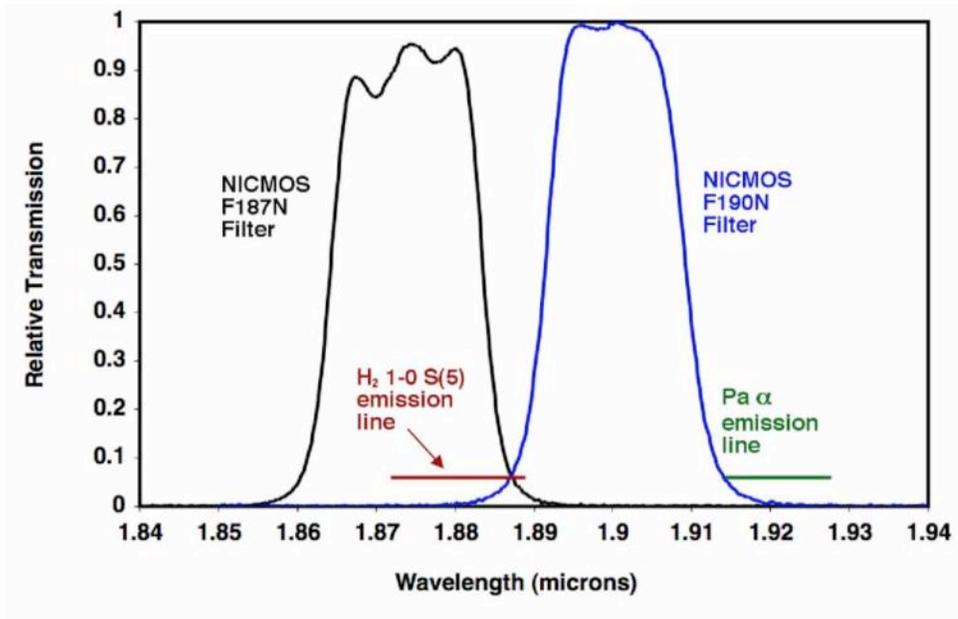

Figure 1c

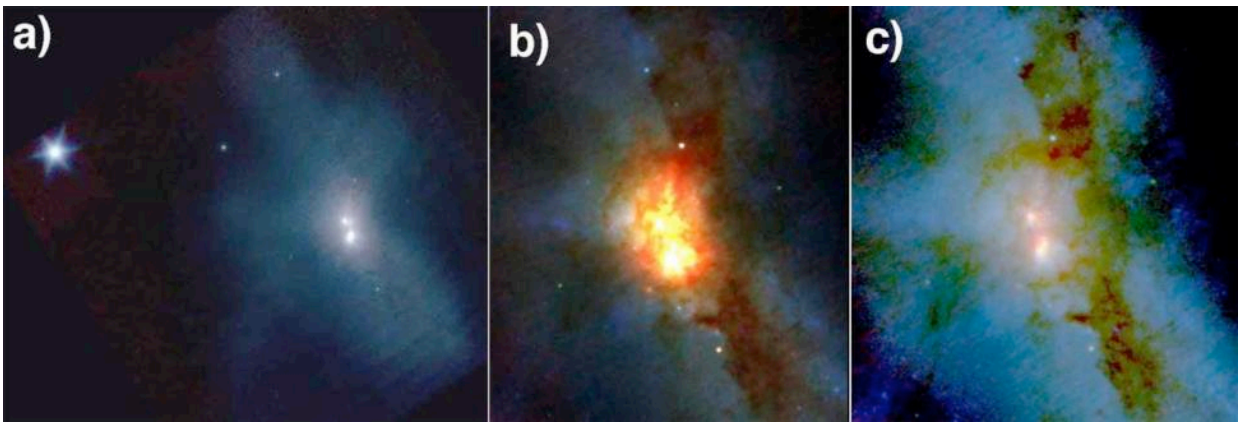

Figure 2

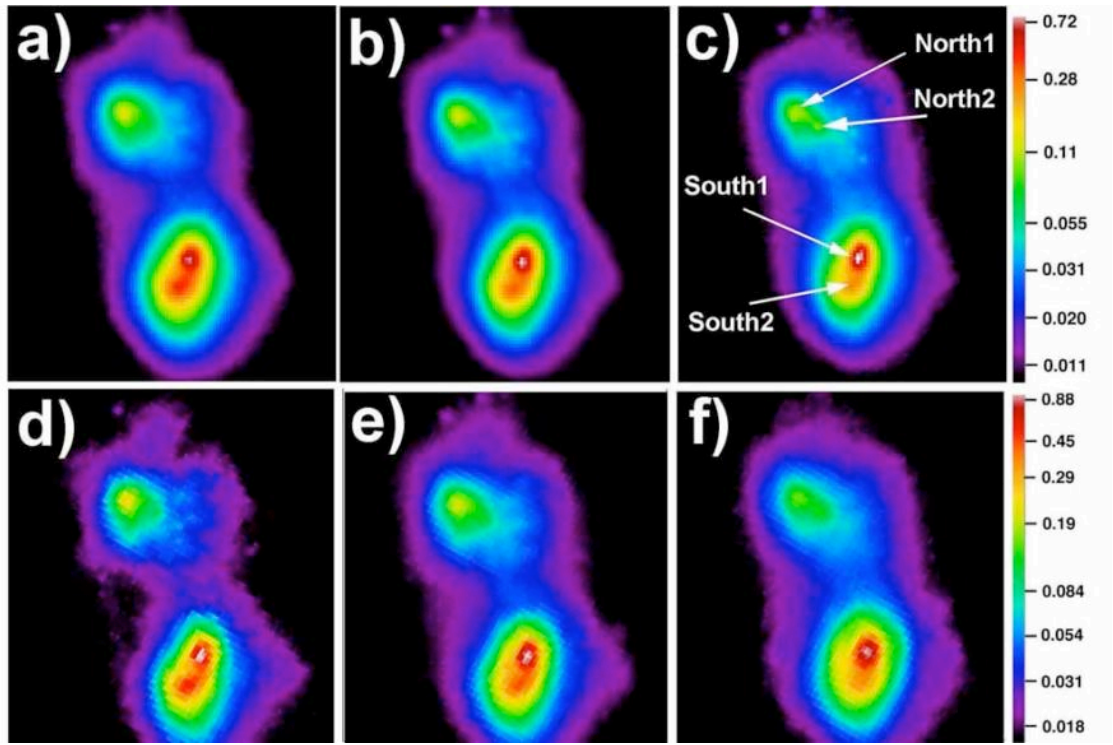

Figure 3

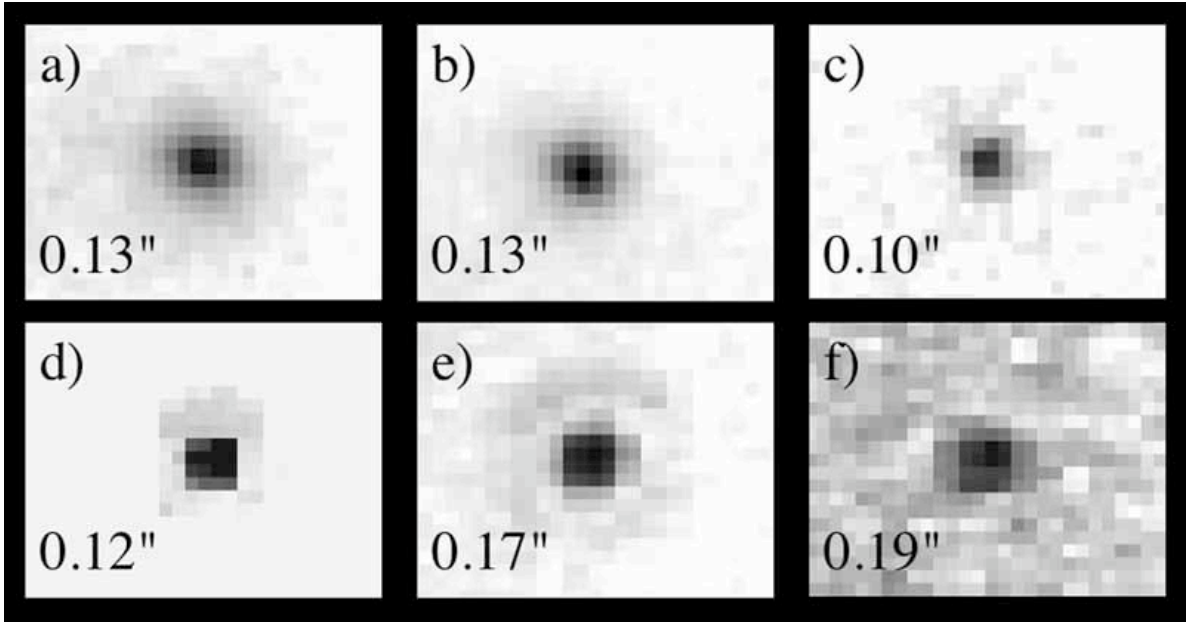

Figure 4

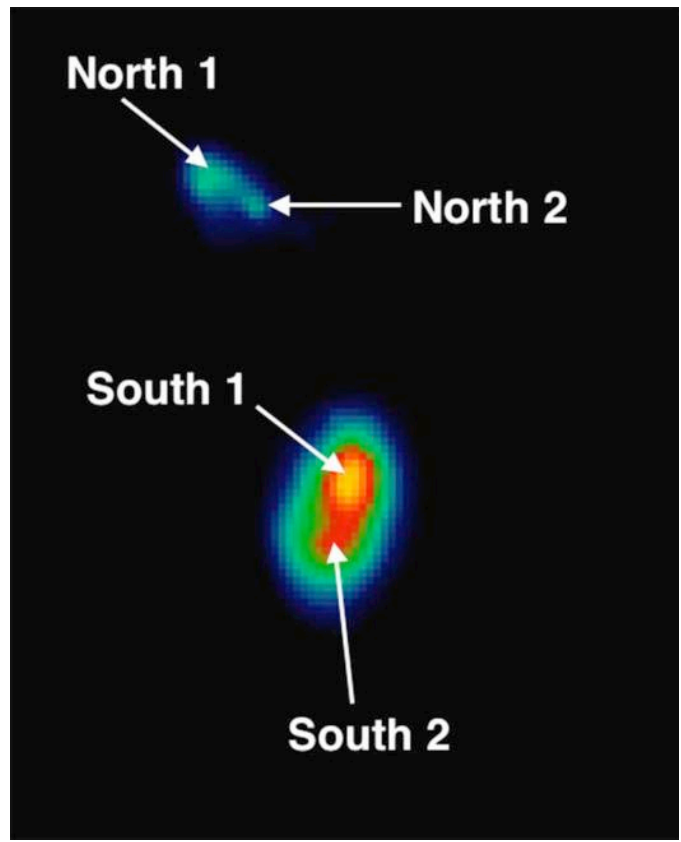

Figure 5

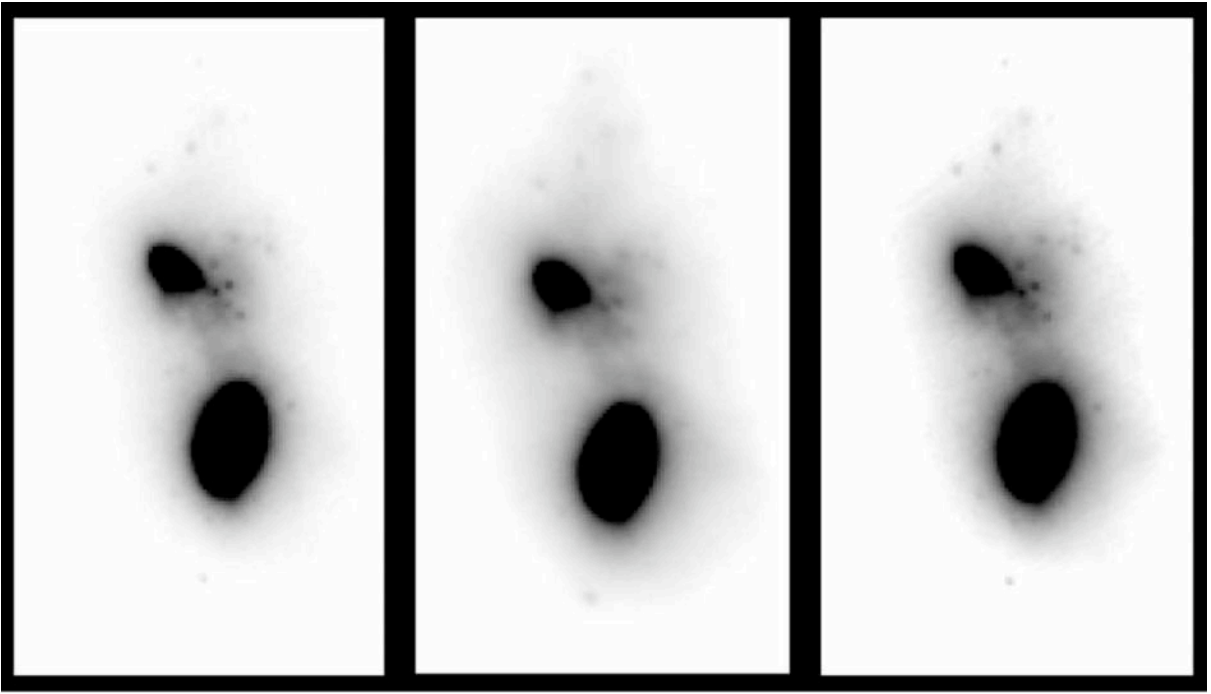

Figure 6

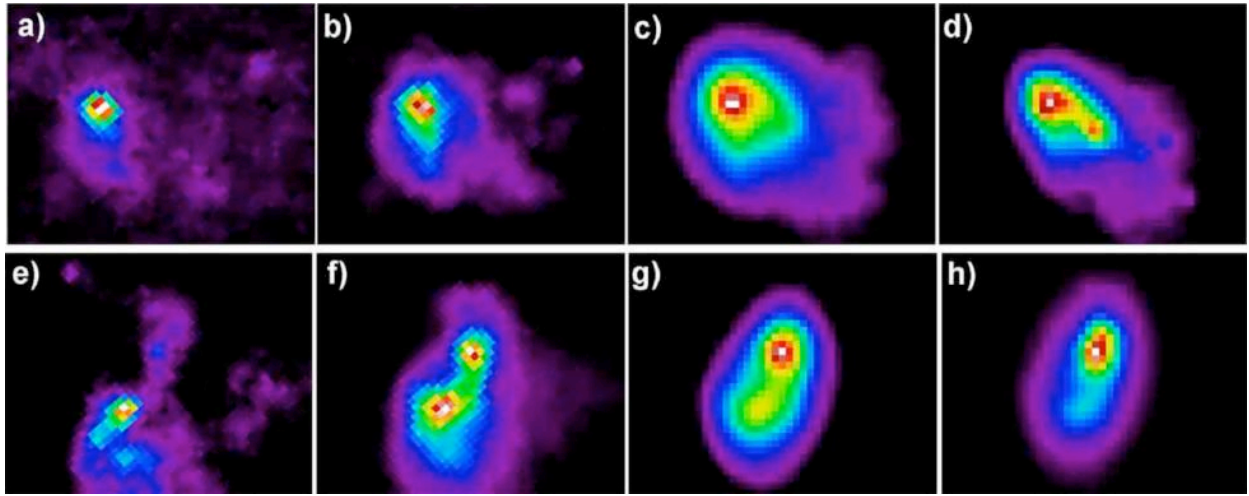

Figure 7

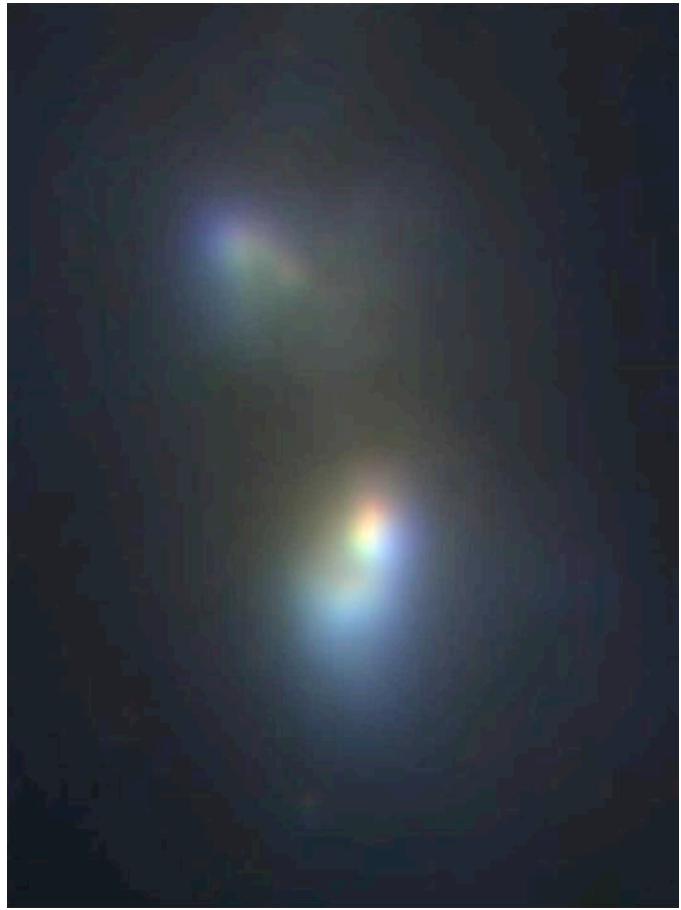

Figure 8

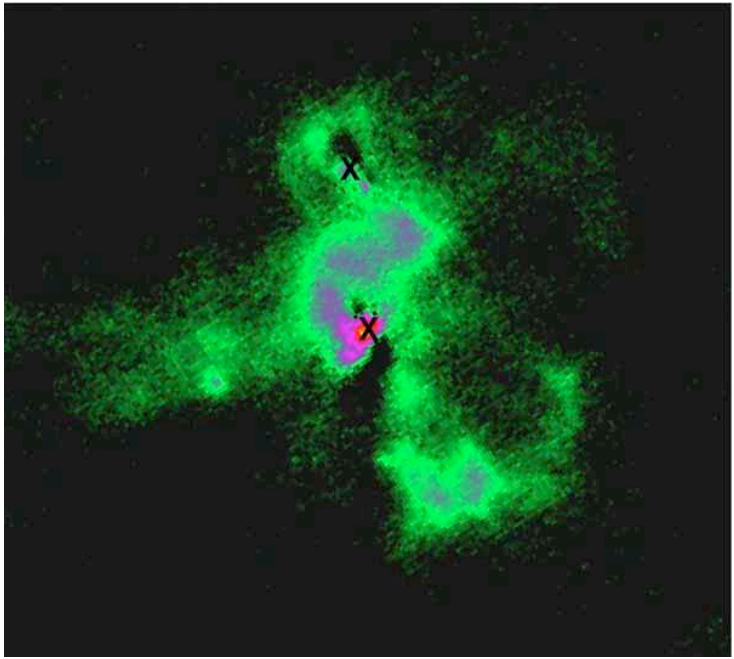

Figure 9a

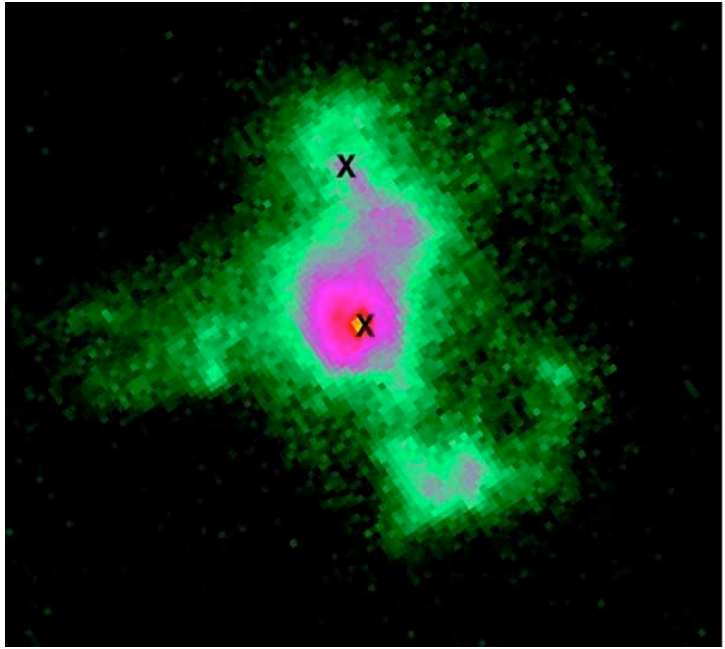

Figure 9b

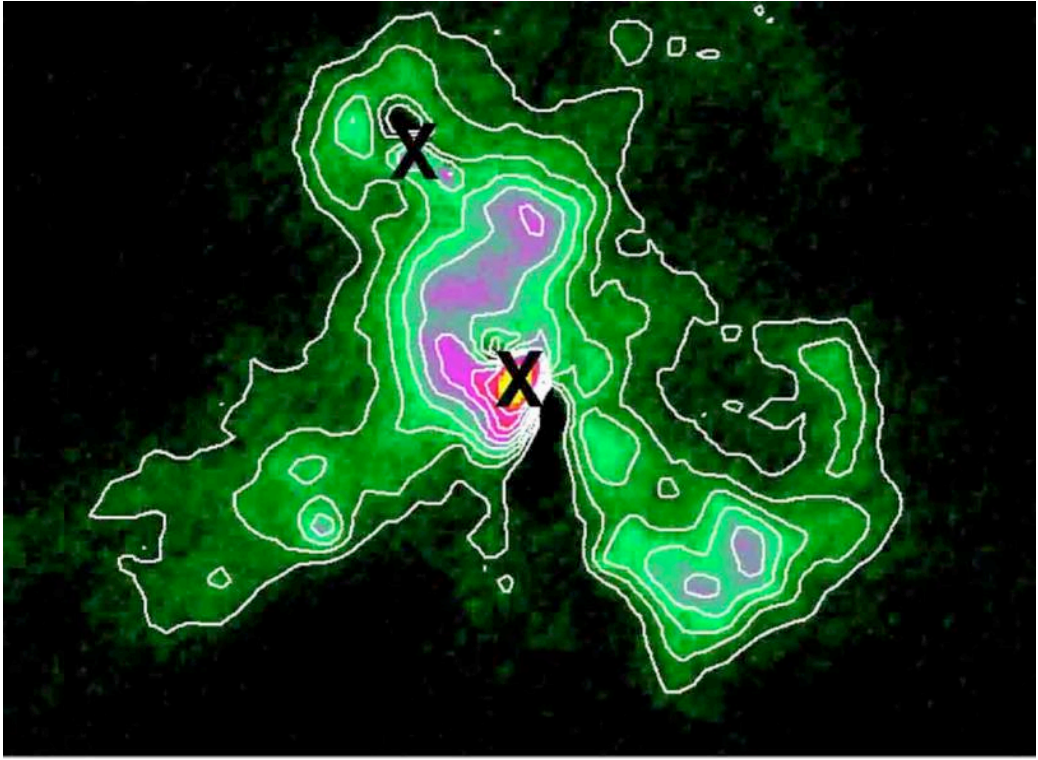

Figure 10

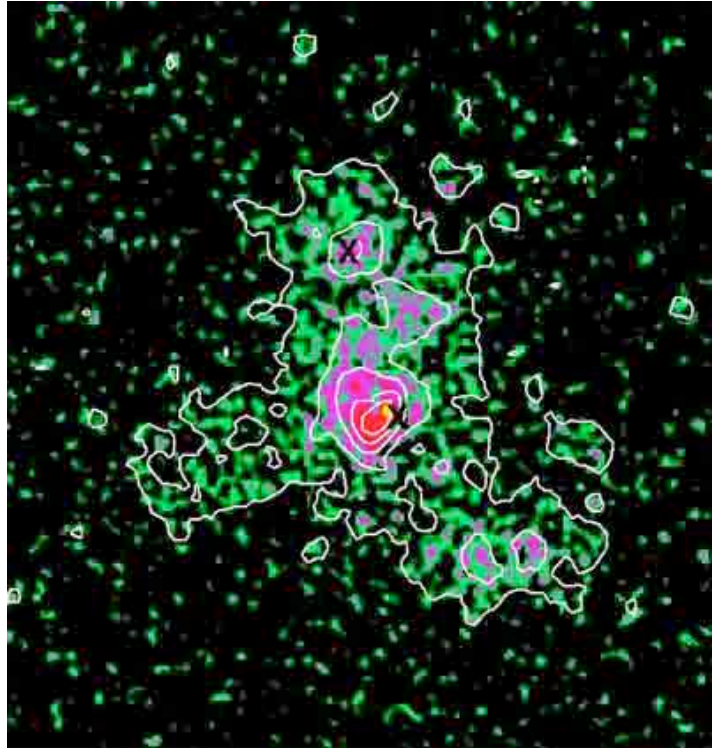

Figure 11

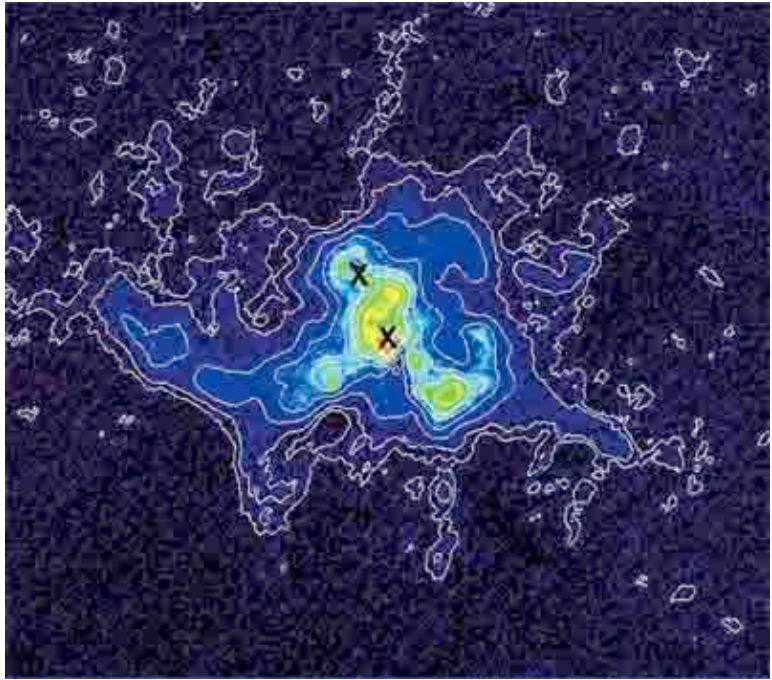

Figure 12a

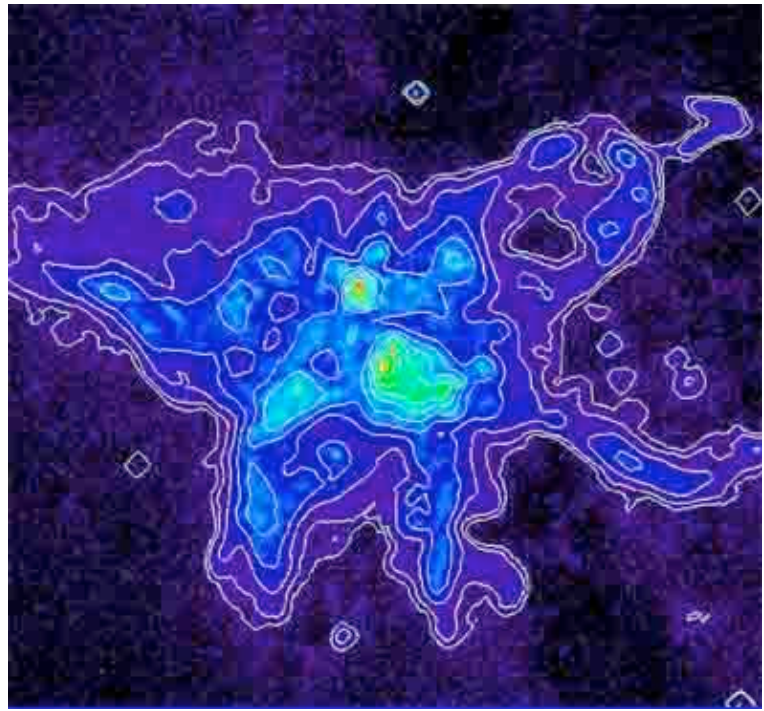

Figure 12b

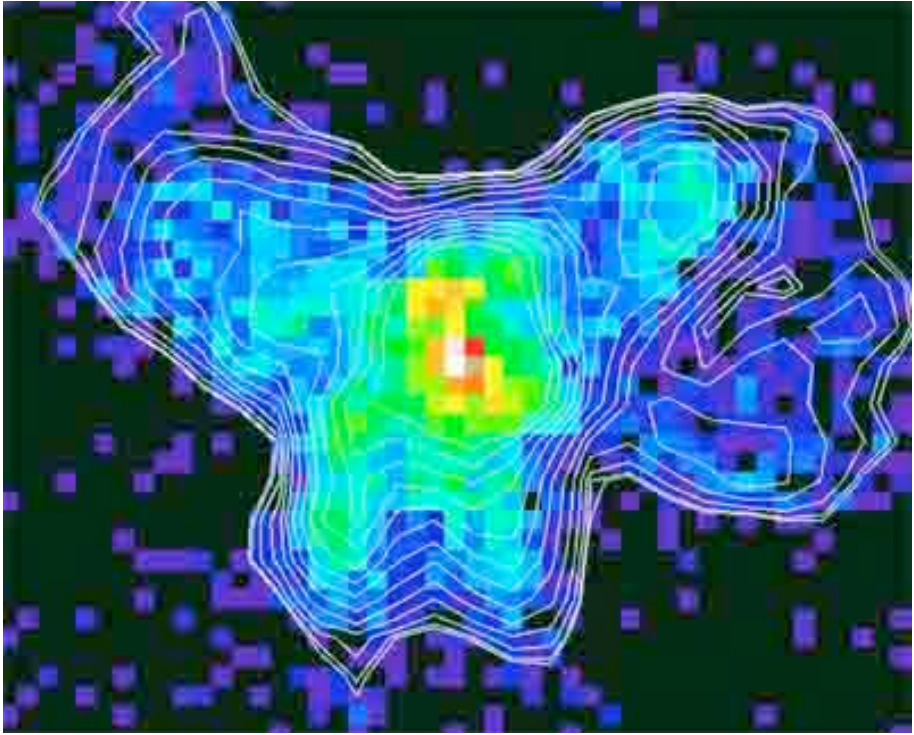

Figure 13